\documentstyle[12pt,a4,subeq,axodraw]{article}

\newcommand\vek[1]{\mbox{\rmfamily\bfseries\itshape#1}}
\newcommand\vekexp[1]{\mbox{\scriptsize\rmfamily\bfseries\itshape#1}}
\def\be{\begin{equation}}
\def\ee{\end{equation}}
\def\ni{\noindent}
\def\e{{\rm e}}
\def\rd{{\rm d}}
\def\bR{{\bf R}}
\def\bF{{\widetilde F}}
\def\vQ{{\vek Q}}

\begin{document}

\centerline{{\bf The Sixth-Moment Sum Rule For the Pair Correlations of}} 
\centerline{{\bf the Two-Dimensional One-Component Plasma: Exact Result}}
\vskip 0.5truecm
\centerline{P. Kalinay, P. Marko\v s, L. \v Samaj and I. Trav\v enec} 
\vskip 0.5truecm
\centerline{Institute of Physics, Slovak Academy of Sciences,} 
\centerline{D\' ubravsk\' a cesta 9, 842 28 Bratislava, Slovakia}
\vskip 0.5truecm
\ni The system under consideration is a two-dimensional one-component
plasma in fluid regime, at density $n$ and at arbitrary coupling
$\Gamma = \beta e^2$ ($e=$ unit charge, $\beta=$ inverse temperature).
The Helmholtz free energy of the model, as the generating functional
for the direct pair correlation $c$, is treated in terms of a
convergent renormalized Mayer diagrammatic expansion in density.
Using specific topological transformations within the bond-renormalized
Mayer expansion we prove that the nonzero contributions to the
regular part of the Fourier component of $c$ up to the $k^2$-term
originate exclusively from the ring diagrams (unable to undertake
the bond-renormalization procedure) of the Helmholtz free energy.
In particular, 
$\hat c({\vek k}) = -\Gamma / k^2 + \Gamma / (8\pi n)
- k^2 / [96 (\pi n)^2] + O(k^4) .$
This result fixes via the Ornstein-Zernike relation, besides 
the well-known zeroth-, second- and fourth-moment sum rules, the
new sixth-moment condition for the truncated pair correlation $h$,
$n (\pi \Gamma n / 2)^3 \int r^6 h({\vek r}) \rd {\vek r}
= 3 (\Gamma -6)(8-3\Gamma)/4.$


\vskip 0.5truecm



\noindent {\bf 1. INTRODUCTION}
\medskip

\noindent Coulomb plasmas are the model systems for studying the effect
of long-range interparticle interactions on statistics of classical 
lattice and continuous fluids.
It was observed that, in arbitrary dimension, the long-range tail of
the Coulomb potential gives rise to exact constraints, sum rules, for
truncated particle correlations (for an exhausting review, see Ref. 1),
namely the zeroth- and second-moment conditions$^{(2,3)}$.

The concentration on two dimensions (2d) with logarithmic interparticle
interactions and on the one-component plasma (OCP), i.e. the continuous
system of charged particles embedded in a spatially uniform background, 
brings some physical peculiarities and relevant mathematical simplifications
providing additional exact information about the system, like:
  
\ni -- A formal relationship to the fractional quantum Hall effect$^{(4)}$;

\ni -- An experimental evidence for the Wigner crystallization at low
temperatures$^{(5)}$;

\ni -- The dependence of the statistics on the only parameter - coupling
constant $\Gamma \sim 1/{\rm temperature}$ (the charge density scales
appropriately the distance);

\ni -- The availability of the equation of state$^{(6)}$;

\ni -- The mapping to free fermions at special coupling $\Gamma =2$$^{(7)}$
(for various sample's geometries, see review 8) characterized by a
pure Gaussian decay of the truncated pair correlation $h$; the evaluation
of the leading term of the $(\Gamma -2)$-expansion of $h$,
indicating the change from monotonic to oscillatory behavior just at
$\Gamma =2$$^{(7)}$;

\ni -- The rigorous derivation of the weak-coupling $\Gamma \to 0$
Debye-H\" uckel limit$^{(9)}$, and the systematic $\Gamma$-expansion
of $h$ in terms of a renormalized Mayer diagrammatic 
expansion$^{(10,11)}$;

\ni -- The shift of the compressibility equation to the fourth-moment
condition$^{(12-14)}$ under the assumption of cluster conditions;

\ni -- A symmetry of thermodynamic quantities with respect to a complex 
transformation of particle coordinates$^{(15)}$ for arbitrary coupling
$\Gamma$, implying a functional relation among the pair correlations. 
The last is equivalent to an infinite sequence of sum rules relating the 
coefficient of the short-distance expansion of two-particle correlations 
(the lowest level of the sequence was derived by Jancovici$^{(16)}$).
The generalization of the symmetry to multi-particle densities, possessing 
specific invariant structure, was given in Ref. 17.

\ni -- The suggestion that, at arbitrary $\Gamma$, the 2d OCP is in 
the critical state$^{(18,19)}$ in terms of the induced electrical-field
correlations (but not the particle correlations).
The free energy is therefore supposed to exhibit finite-size correction
predicted by the conformal-invariance theory, as was verified
by the rigorous finite-size treatment of the $\Gamma =2$ case and also 
numerically$^{(20)}$, by using exact finite-size techniques$^{(4,21-23)}$, 
for coupling strengths $\Gamma=4$ and $\Gamma=6$. 

The present paper is devoted to a rigorous derivation of the new sixth-moment
sum rule for the truncated pair correlation $h$ of the 2d OCP.
The mathematical basis comes from the convergent bond-renormalized Mayer
expansion in density$^{(11)}$.
Within a specific classification of the diagrams in the renormalized
format for the Helmholtz free energy, the functional generator for
the direct pair correlation $c$, we prove that the regular part of the
Fourier component of $c$ is determined up to the $k^2$-term solely
by the (unrenormalized) ring diagrams of the generating free energy.
This result implies via the Ornstein-Zernike (OZ) relation, besides
the known zeroth-, second-$^{(2,3)}$ and fourth-moment$^{(12-14)}$
sum rules, the explicit formula for the sixth moment of $h$.

The paper is outlined as follows:

In section 2, we recapitulate briefly the ordinary Mayer diagrammatic 
representation in density for $h$ and $c$ pair correlations and
the (excess) Helmholtz free energy as the generating functional for 
both of them.

Section 3 deals with exactly solvable cases or limits
of the 2d OCP, expressed in terms of $h$-moments.
These involve the momentum sum rules,
the $\Gamma = 2$ coupling together with the leading $(\Gamma -2)$
correction term and the Debye-H\" uckel $\Gamma\to 0$ limit.
Here, we have registered a very important fact.
The three known (appropriately rescaled) zeroth, second and fourth
moments turn out to be $\Gamma$-polynomials of (successively increasing)
{\it finite} order.
By using the renormalized Mayer expansion$^{(11)}$ we were able to compute
for the as-yet-unknown sixth $h$-moment the coefficients to a few lower 
orders of the $\Gamma$-expansion terms around the Debye-H\" uckel limit and
observed, within the range of these low orders, the finite 
$\Gamma$-truncation also for this case.
To our surprise, this finite truncation represented an exact interpolation
between the $\Gamma\to 0$ and $\Gamma=2$ couplings and, moreover,
reproduced correctly the leading $(\Gamma-2)$ correction term$^{(7)}$.
Since the coupling $\Gamma =2$ and its neighborhood do not play any special
role in view of the $\Gamma$-expansion, the above fact was a strong 
indication and motivation for us.

Section 4 describes the formalism of the renormalized Mayer expansion$^{(11)}$.
The renormalization of bond factors consists in a multiple-bond expansion
of Mayer functions and a consequent series elimination of field circles,
resulting in the modified Bessel functions of second kind.
The novelty lies in the classification of diagrams representing the
Helmholtz free energy -- the functional generator of $c$ -- according
to the possibility of performing the series-elimination transformation:
(1) simple unrenormalized bond generates the characteristic singular term
of $c$; (2) all unrenormalized ring diagrams (which cannot undertake the
series-elimination procedure) generate the renormalized ``watermelon'' 
Meeron graph contribution to the regular part of $c$; (3) every other
diagram is expressible with all bonds renormalized and as generator
gives rise to a family of $c$-diagrams: the families do not overlap with
one another and, as units, they exhibit remarkable ``cancellation properties''.

The ``cancellation'' phenomena is the subject of the crucial section 5
where we prove that, regardless of the topology of a separate graph belonging
to class (3), the zeroth and second real-space moments of the $c$-diagrams
family, generated from the underlying graph, vanish.  
The proof of the nullity of the zeroth moment follows from a trivial 
scaling property of the Bessel functions with respect to density.
The proof of the second-moment condition is much more complicated.
Besides the above scaling property it requires to introduce an elimination
procedure for two-coordinated root points generated on renormalized
bonds and to reveal ``hidden zeros'' due to the translational and 
rotational invariance of the infinite system, realized through 
per-partes integration of field-point coordinates.

In Conclusion, after evaluating the only contribution to the Fourier
component of $c$ up to the $k^2$-term, namely that
of the renormalized Meeron graph, we write down by using the OZ relation
the explicit formula for the sixth moment of $h$.
The structure of higher-order moments of $h$ is also discussed.

\vskip 1.5truecm
\noindent {\bf 2. A SKETCH OF THE MAYER EXPANSION IN DENSITY$^{(24)}$}
\medskip

We consider a system of identical pointlike particles in volume $V$ of a
$d$-dimensional space, interacting through pair potential $v$;
$v$ will occur in combination called the Mayer function
\be \label{1}
f(i,j) = \exp [-\beta v(i,j)] -1 
\ee
where $\beta$ is the inverse temperature and, for notational convenience,
a position vector ${\vek r}_i$ is represented simply by $i$.  
In the inverse (density) format, i.e. with the density $n({\vek r}) = \langle
\sum_i \delta({\vek r}-{\vek r}_i)\rangle$ as controlling variable,
the excess free energy $\bar F^{ex}$ is the relevant thermodynamic
potential.
Its Mayer diagrammatic representation reads
\begin{eqnarray} \label{2}
\beta \bar F^{ex} & = & \big\{ {\rm all\ connected\ diagrams\ which\
consist\ of\ } N\ge 2\ {\rm field\ (black)}\nonumber \\
& & n{\rm -circles\ and\ } f{\rm -bonds,\ and\ are\ free\ of\ connecting\
circles} \big\}
\end{eqnarray}
(the removal of a connecting circle disconnects the diagram).
The excess free energy is the generating functional for the truncated
pair correlation
\be \label{3}
h(1,2) = {n_2(1,2) - n(1) n(2) \over n(1) n(2)} 
\ee
with the two-body density $n_2({\vek r},{\vek r}') = \langle
\sum_{i\ne j} \delta({\vek r}-{\vek r}_i) \delta({\vek r}'-{\vek r}_j)\rangle$
and for the direct correlation function $c$, in the sense that
\begin{eqnarray} \label{4,5}
h(1,2) & = & -1 + [1+f(1,2)] \displaystyle{1\over n(1) n(2)} 
\displaystyle{\delta \beta \bar F^{ex} \over \delta f(1,2)} \\
c(1,2) & = & \displaystyle{\delta^2 \beta \bar F^{ex} \over \delta n(1) 
\delta n(2)}
\end{eqnarray}
With regard to (\ref{2}), this implies
\begin{eqnarray} \label{6}
h(1,2) = & \big\{ {\rm all\ connected\ 1,2-rooted\ diagrams\ which\
consist\ of\ field} \nonumber \\
& n{\rm -circles\ and\ } f{\rm -bonds,\ and\ are\ free\ of\ articulation\
circles} \big\}
\end{eqnarray}
[the removal of an articulation circle disconnects the diagram into
two or more components, of which at least one contains no root (white)
circle];
\begin{eqnarray} \label{7}
c(1,2) = & \big\{ {\rm all\ connected\ 1,2-rooted\ diagrams\ which\
consist\ of\ field} \nonumber \\
& n{\rm -circles\ and\ } f{\rm -bonds,\ and\ are\ free\ of\ connecting\
circles} \big\}
\end{eqnarray}
The link between $h$ and $c$ is established in terms of the
OZ relation
\be \label{8}
h(1,2) = c(1,2) + \int c(1,3) n(3) h(3,2) \rd 3 
\ee
If the system is infinite $(V\to\infty)$, homogeneous, $n(1) = n$,
and both isotropic and translationally invariant, $h(1,2) = h(\vert 1-2\vert)$,
$c(1,2) = c(\vert 1-2\vert)$,
it is useful to introduce the Fourier components
\begin{subeqnarray} \label{9}
f({\vek r}) & = & \displaystyle{1\over (2\pi)^{d/2}} \int \exp(i{\vek k}\cdot 
{\vek r}) \hat f({\vek k}) \rd {\vek k} \\
\hat f({\vek k}) & = & \displaystyle{1\over (2\pi)^{d/2}} \int \exp(-i{\vek k}
\cdot {\vek r}) f({\vek r}) \rd {\vek r}
\end{subeqnarray}
Especially, in $d=2$ dimensions,
\begin{eqnarray} \label{10}
\hat f({\vek k}) & = & \displaystyle{\int} r f(r) J_0(k r) \rd r
\nonumber \\
& = & \sum_{j=0}^{\infty} \displaystyle{(-1)^j\over (j!)^2} 
\left( \displaystyle{k^2\over 4} \right)^j 
\displaystyle{1\over 2\pi} \int r^{2j} f(r) \rd {\vek r}
\end{eqnarray}
with $J_0$ being the ordinary Bessel function.
In the Fourier space, the OZ relation (\ref{8}) takes the form
$$\hat h(k) = \hat c(k) + (2\pi)^{d/2} n \hat c(k) \hat h(k) \eqno(8')$$ 
where $k = \vert {\vek k} \vert$.
\newpage

\noindent {\bf 3. MOTIVATION}
\medskip

The classical OCP is a system of particles of charge $e$ embedded in a
spatially uniform neutralizing background.
In $d=2$ dimensions, the Coulomb interaction energy is given by
\begin{subeqnarray} \label{11}
-\beta v(1,2) & = & \Gamma \ln \vert 1-2 \vert \\
-\beta \hat v({\vek k}) & = & - \Gamma / k^2
\end{subeqnarray}
with $\Gamma = \beta e^2$ being the coupling constant.
We will concentrate on the thermodynamic limit of the fluid regime
with constant density $n(i) = n$ and use the notation
\be \label{12}
I_{2j} = \int r^{2j} h({\vek r}) \rd {\vek r} 
\ee
for the moments of the truncated pair correlation.

Let us first summarize exactly solvable cases of the model.
In the weak coupling $\Gamma\to 0$ limit, $h$ displays the
Debye-H\" uckel screening$^{(9)}$
\be \label{13}
h(r;\Gamma\to 0) \simeq - \Gamma K_0(r\sqrt{2\pi \Gamma n}) 
\ee
with $K_0$ the modified Bessel function of second kind.
Consequently,
\be \label{14}
\lim_{\Gamma\to 0} n \left( {\pi \Gamma n \over 2}\right)^j 
I_{2j}(\Gamma) = - (j!)^2 
\ee
At $\Gamma = 2$, the mapping onto free fermions provides a pure 
Gaussian form of $h$$^{(7)}$,
\be \label{15}
h(r;\Gamma=2) = - \exp (-\pi n r^2) 
\ee
implying
\be \label{16}
n (\pi n)^j I_{2j}(\Gamma =2) = - j!
\ee
The leading-order of the series expansion around $\Gamma =2$ results in$^{(7)}$
\begin{subeqnarray} \label{17}
h(r;\Gamma) & = & h(r;\Gamma=2) + (\Gamma-2) \delta h(r) + \ldots \\
\delta h(r) & = & {\rm Ei}(-\pi n r^2) - {1\over 2} {\rm Ei}(-\pi n r^2/2)
\nonumber  \\
& & + \exp(-\pi n r^2) \left\{ {1\over 2} {\rm Ei}(\pi n r^2/2)
- [\ln(\pi n r^2) + C] \right\} 
\end{subeqnarray}
where $C$ is Euler's constant and Ei the exponential-integral function.
From (\ref{17}) one gets after some algebra
\be \label{18}
n \left( {\pi \Gamma n \over 2} \right)^j I_{2j}(\Gamma) = - j!
+(\Gamma-2)j! \left( \sum_{k=0}^j {2^k-1\over k+1} - {j\over 2} \right)
+ O[(\Gamma-2)^2] 
\ee

The long-range tail of the Coulomb potential gives rise to exact constraints 
(sum rules) for the moments of the truncated two-body correlation,
like the zeroth-moment (perfect screening) condition
\begin{subequations} \label{19} 
\be
n I_0  =  - 1 
\ee
the second-moment (Stillinger-Lovett) condition$^{(2,3)}$
\be
n \left( {\pi \Gamma n \over 2} \right) I_2 = -1 
\ee
the fourth-moment (compressibility) condition$^{(12-14)}$
\be
n \left( {\pi \Gamma n \over 2} \right)^2 I_4 = -4 + \Gamma
\ee
\end{subequations}
Note that the sum rules are consistent with exact formulae
(\ref{14}), (\ref{18}).

The expressions for the rescaled moments (\ref{19}) correspond to
finite truncations of their $\Gamma$-expansion around $\Gamma=0$.
The $\Gamma$-expansion technique of $h$$^{(11)}$, explained and extended
in the next section, enables one to evaluate systematically
the coefficients of the $\Gamma$-expansion also for higher moments.
For the sixth moment of $h$, we were able to attain with a little effort
the third order of $\Gamma$, with the result
\be \label{20}
n \left( {\pi \Gamma n \over 2} \right)^3 I_6 = -36 +
{39\over 2} \Gamma - {9\over 4} \Gamma^2 + 0 \times \Gamma^3
+ O(\Gamma^4) 
\ee
The appearance of the zero coefficient to the $\Gamma^3$ power
indicates the possibility of a finite $\Gamma$-truncation for $I_6$, too.
This hypothesis is strongly supported by the fact that the truncation
of (\ref{20}) at the $\Gamma^2$-term interpolates correctly from 
the $\Gamma\to 0$ limit [relation (\ref{14})] to the $\Gamma=2$ coupling 
in the sense that 
the rescaled moment $I_6$ satisfies (\ref{18}), i.e., the rhs of 
(\ref{20}) acquires the exact value $(-6)$ at $\Gamma=2$ and exhibits 
the exact prefactor $21/2$ to the leading $(\Gamma-2)$ correction term.
These facts were our primary motivation for proving rigorously
the conjecture of the finite $\Gamma$-series-truncation of $I_6$.

It turns out to be more convenient to formulate the above sum rules 
in terms of the small-${\vek k}$ expansion of the Fourier component
of the direct correlation, $\hat c({\vek k})$.
Inserting (\ref{19}) together with the suggested truncation of the sixth
moment at the $\Gamma^2$-term (\ref{20}) into the representation 
(\ref{10}) for $\hat h({\vek k})$, the OZ relation (8') implies
the expected form
\be \label{21}
\hat c({\vek k}) = - {\Gamma \over k^2} + {\Gamma \over 8 \pi n}
- {k^2 \over 96 (\pi n)^2} + O(k^4) 
\ee
Note the characteristic singular leading term of $\hat c({\vek k})$,
$-\beta \hat v({\vek k}) = - \Gamma/k^2$, succeeded by the regular
$k^2$-series expansion part: its knowledge up to the $k^{2j}$ term
determines $\hat h({\vek k})$ up to the $k^{2(j+2)}$ term, or
equivalently, the real-space $h$-moments (\ref{12}) up to $I_{2(j+2)}$.

\vskip 2truecm

\noindent {\bf 4. RENORMALIZED MAYER EXPANSION}
\medskip

The Mayer function $f$ can be expanded in the inverse temperature 
$\beta$ as follows
\be \label{22}
f(1,2) = -\beta v(1,2) + {1\over 2!} [-\beta v(1,2)]^2
+ {1\over 3!} [-\beta v(1,2)]^3 + \ldots 
\ee
Graphically,
\vspace{5pt} \hfill\\
\noindent
\begin{picture}(55,40)(0,7)
    \Line(0,10)(50,10)
    \BCirc(0,10){2.5} \BCirc(50,10){2.5}
    \Text(0,0)[]{1} \Text(50,0)[]{2}
    \Text(25,23)[]{$f$}
\end{picture} $\ =\ \ $
\begin{picture}(55,20)(0,7)
    \DashLine(0,10)(50,10){7}
    \BCirc(0,10){2.5} \BCirc(50,10){2.5}
    \Text(0,0)[]{1} \Text(50,0)[]{2}
    \Text(24,23)[]{$-\beta v$}
\end{picture} $\ +\ \ $
\begin{picture}(55,25)(0,7)
    \DashCArc(25,-14)(34,45,135){5}
    \DashCArc(25,34)(34,225,315){5}
    \BCirc(0,10){2.5} \BCirc(50,10){2.5}
    \Text(0,0)[]{1} \Text(50,0)[]{2}
\end{picture} $\ +\ \ $
\begin{picture}(55,40)(0,7)
    \DashCArc(25,-14)(34,45,135){5}
    \DashCArc(25,34)(34,225,315){5}
    \DashLine(0,10)(50,10){5}
    \BCirc(0,10){2.5} \BCirc(50,10){2.5}
    \Text(0,0)[]{1} \Text(50,0)[]{2}
\end{picture} $\ +\ \  ...$
\hfill {(22')}
\vskip1cm

\noindent 
where the factor $1/({\rm number\ of\ interaction\ lines})!$
is automatically involved in each diagram.
Let us perform the above $f$-decomposition within 
the diagrammatic representation (\ref{2}) of $\beta \bar F^{ex}$:

\noindent
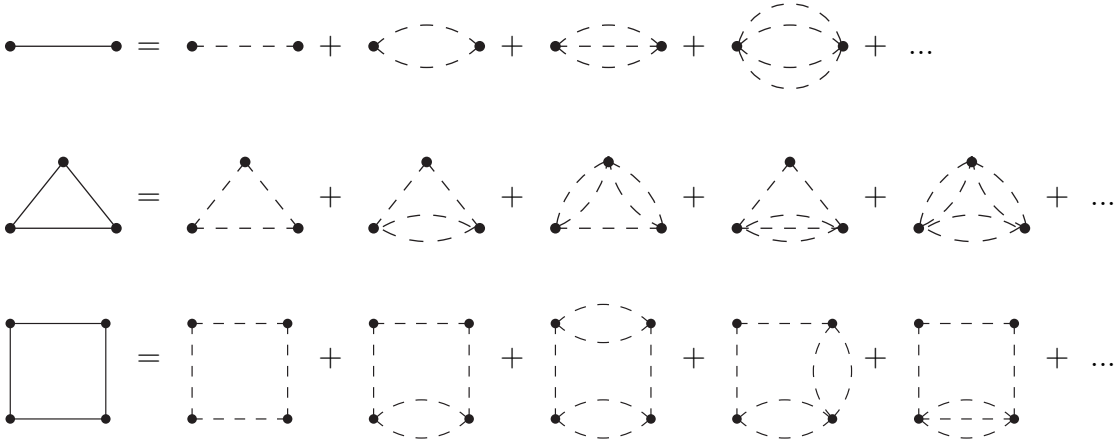
\begin{picture}(40,40)(0,7)
    \Line(0,10)(40,10)
    \Vertex(0,10){2} \Vertex(40,10){2}
\end{picture} $\ =\ \ $
\begin{picture}(40,40)(0,7)
    \DashLine(0,10)(40,10){5}
    \Vertex(0,10){2} \Vertex(40,10){2}
\end{picture} $\ +\ \ $
\begin{picture}(40,40)(0,7)
    \DashCArc(20,-10)(28,45,135){5}
    \DashCArc(20,30)(28,225,315){5}
    \Vertex(0,10){2} \Vertex(40,10){2}
\end{picture} $\ +\ \ $
\begin{picture}(40,40)(0,7)
    \DashCArc(20,-10)(28,45,135){5}
    \DashCArc(20,30)(28,225,315){5}
    \DashLine(0,10)(40,10){5}
    \Vertex(0,10){2} \Vertex(40,10){2}
\end{picture} $\ +\ \ $
\begin{picture}(40,40)(0,7)
    \DashCArc(20,-10)(28,45,135){5}
    \DashCArc(20,30)(28,225,315){5}
    \DashCArc(20,6)(20,15,165){5}
    \DashCArc(20,14)(20,195,345){5}
    \Vertex(0,10){2} \Vertex(40,10){2}
\end{picture} $\ +\ \  ...$
\hfill \\
\vskip0.05truecm

\noindent
\begin{picture}(40,40)(0,19)
    \Line(0,10)(40,10)
    \Line(0,10)(20,35)
    \Line(20,35)(40,10)
    \Vertex(0,10){2} \Vertex(40,10){2} \Vertex(20,35){2}
\end{picture} $\ =\ \ $
\begin{picture}(40,20)(0,19)
    \DashLine(0,10)(40,10){5}
    \DashLine(0,10)(20,35){5}
    \DashLine(20,35)(40,10){5}
    \Vertex(0,10){2} \Vertex(40,10){2} \Vertex(20,35){2}
\end{picture} $\ +\ \ $
\begin{picture}(40,20)(0,19)
    \DashLine(0,10)(20,35){5}
    \DashLine(20,35)(40,10){5}
    \Vertex(0,10){2} \Vertex(40,10){2} \Vertex(20,35){2}
    \DashCArc(20,-18)(33,60,120){5}
    \DashCArc(20,38)(33,240,305){5}
\end{picture} $\ +\ \ $
\begin{picture}(40,20)(0,19)
    \DashCArc(32,6)(32,115,170){5}
    \DashCArc(-12,40)(32,295,355){5}
    \DashCArc(9,5)(32,10,70){5}
    \DashCArc(52,40)(32,185,250){5}
    \DashLine(0,10)(40,10){5}
    \Vertex(0,10){2} \Vertex(40,10){2} \Vertex(20,35){2}
\end{picture} $\ +\ \ $
\begin{picture}(40,20)(0,19)
    \DashLine(0,10)(20,35){5}
    \DashLine(20,35)(40,10){5}
    \DashLine(0,10)(40,10){5}
    \Vertex(0,10){2} \Vertex(40,10){2} \Vertex(20,35){2}
    \DashCArc(20,-18)(33,60,120){5}
    \DashCArc(20,38)(33,240,305){5}
\end{picture} $\ +\ \ $
\begin{picture}(40,20)(0,19)
    \DashCArc(32,6)(32,115,170){5}
    \DashCArc(-12,40)(32,295,355){5}
    \DashCArc(9,5)(32,10,70){5}
    \DashCArc(52,40)(32,185,250){5}
    \DashCArc(20,-18)(33,60,120){5}
    \DashCArc(20,38)(33,240,305){5}
    \Vertex(0,10){2} \Vertex(40,10){2} \Vertex(20,35){2}
\end{picture} $\ +\ \  ...$
\hfill \\
\vskip0.2truecm
\SetScale{0.9}
\noindent
\begin{picture}(40,40)(0,20)
    \Line(0,0)(40,0)
    \Line(0,0)(0,40)
    \Line(0,40)(40,40)
    \Line(40,0)(40,40)
    \Vertex(0,0){2} \Vertex(40,0){2} \Vertex(0,40){2} \Vertex(40,40){2}
\end{picture} $\ =\ \ $
\begin{picture}(40,40)(0,20)
    \DashLine(0,0)(40,0){5}
    \DashLine(0,0)(0,40){5}
    \DashLine(0,40)(40,40){5}
    \DashLine(40,0)(40,40){5}
    \Vertex(0,0){2} \Vertex(40,0){2} \Vertex(0,40){2} \Vertex(40,40){2}
\end{picture} $\ +\ \ $
\begin{picture}(40,40)(0,20)
    \DashLine(0,0)(0,40){5}
    \DashLine(0,40)(40,40){5}
    \DashLine(40,0)(40,40){5}
    \Vertex(0,0){2} \Vertex(40,0){2} \Vertex(0,40){2} \Vertex(40,40){2}
    \DashCArc(20,-20)(28,45,135){5}
    \DashCArc(20,20)(28,225,315){5}
\end{picture} $\ +\ \ $
\begin{picture}(40,40)(0,20)
    \DashLine(0,0)(0,40){5}
    \DashLine(40,0)(40,40){5}
    \Vertex(0,0){2} \Vertex(40,0){2} \Vertex(0,40){2} \Vertex(40,40){2}
    \DashCArc(20,-20)(28,45,135){5}
    \DashCArc(20,20)(28,225,315){5}
    \DashCArc(20,20)(28,45,135){5}
    \DashCArc(20,60)(28,225,315){5}
\end{picture} $\ +\ \ $
\begin{picture}(40,40)(0,20)
    \DashLine(0,0)(0,40){5}
    \DashLine(0,40)(40,40){5}
    \Vertex(0,0){2} \Vertex(40,0){2} \Vertex(0,40){2} \Vertex(40,40){2}
    \DashCArc(20,-20)(28,45,135){5}
    \DashCArc(20,20)(28,225,315){5}
    \DashCArc(20,20)(28,315,45){5}
    \DashCArc(60,20)(28,135,225){5}
\end{picture} $\ +\ \ $
\begin{picture}(40,40)(0,20)
    \DashLine(0,0)(0,40){5}
    \DashLine(0,40)(40,40){5}
    \DashLine(40,0)(40,40){5}
    \DashLine(0,0)(40,0){5}
    \Vertex(0,0){2} \Vertex(40,0){2} \Vertex(0,40){2} \Vertex(40,40){2}
    \DashCArc(20,-20)(28,45,135){5}
    \DashCArc(20,20)(28,225,315){5}
\end{picture} $\ +\ \ ...$

\hfill \\
\SetScale{1}
\vskip0.1cm

\noindent etc.
If there are only one- or two-coordinated field circles in a graph
(under coordination of a vertex we mean the number of bonds met at
this vertex), we do nothing.
If there are some three- or more-coordinated field circles in a graph,
we can eliminate all two-coordinated field circles by a series
transformation and arrive at a connected graph of field circles
of coordination $\ge 3$, called skeleton.
Grouping the diagrams which are reduced to the same skeleton after series
elimination, the bonds connecting skeleton field circles become
dressed according to

\noindent
\begin{eqnarray} \label{23} 
K(1,2) & = &
\begin{picture}(32,20)(0,7)
    \DashLine(0,10)(32,10){5}
    \BCirc(0,10){2.5} \BCirc(32,10){2.5}
    \Text(0,0)[]{1} \Text(32,0)[]{2}
\end{picture} \ +\ \
\begin{picture}(64,20)(0,7)
    \DashLine(0,10)(32,10){5}
    \DashLine(32,10)(64,10){5}
    \BCirc(0,10){2.5} \BCirc(64,10){2.5}
    \Vertex(32,10){2.2}
    \Text(0,0)[]{1} \Text(64,0)[]{2}
\end{picture} \ +\ \
\begin{picture}(96,20)(0,7)
    \DashLine(0,10)(32,10){5}
    \DashLine(32,10)(64,10){5}
    \DashLine(64,10)(96,10){5}
    \BCirc(0,10){2.5} \BCirc(96,10){2.5}
    \Vertex(32,10){2.2} \Vertex(64,10){2.2}
    \Text(0,0)[]{1} \Text(96,0)[]{2}
\end{picture} \ + \ldots \nonumber \\
& = & 
\begin{picture}(32,20)(0,7)
    \Photon(0,10)(32,10){1}{7}
    \BCirc(0,10){2.5} \BCirc(32,10){2.5}
    \Text(0,0)[]{1} \Text(32,0)[]{2}
\end{picture}
\end{eqnarray}
\noindent Equivalently,
$$K(1,2) = [-\beta u(1,2)] + \int [-\beta u(1,3)] n(3) K(3,2) \rd 3 
\eqno(23')$$
or, in the case of an infinite homogeneous fluid,
$$\hat K({\vek k}) = [-\beta \hat u({\vek k})] + (2\pi)^{d/2} n [-\beta
\hat u({\vek k})] \hat K({\vek k}) \eqno(23'')$$
The procedure of bond renormalization thus implies
\begin{subequations} \label{24}
\be
\beta \bar F^{ex}[n] = \ \  
\begin{picture}(50,20)(0,7)
    \DashLine(0,10)(40,10){5}
    \Vertex(0,10){2.2} \Vertex(40,10){2.2}
\end{picture}
 + \ D_0[n] + \ \sum_{s=1}^{\infty}\ D_s[n] 
\ee
where $D_0$ is the sum of all unrenormalized ring diagrams (which
cannot undertake the renormalization procedure)
\be
D_0 = 
\begin{picture}(40,40)(0,7)
    \DashCArc(20,-10)(28,45,135){5}
    \DashCArc(20,30)(28,225,315){5}
    \Vertex(0,10){2} \Vertex(40,10){2}
\end{picture} \ \ +\ \ 
\begin{picture}(40,20)(0,19)
    \DashLine(0,10)(40,10){5}
    \DashLine(0,10)(20,37){5}
    \DashLine(20,37)(40,10){5}
    \Vertex(0,10){2} \Vertex(40,10){2} \Vertex(20,37){2}
\end{picture} \ \ +\ \ 
\begin{picture}(30,30)(0,10)
    \DashLine(0,0)(30,0){5}
    \DashLine(0,0)(0,30){5}
    \DashLine(0,30)(30,30){5}
    \DashLine(30,0)(30,30){5}
    \Vertex(0,0){2} \Vertex(30,0){2} \Vertex(0,30){2} \Vertex(30,30){2}
\end{picture} \ \ +\ \ \ldots 
\ee
and
\begin{eqnarray}
\sum_{s=1}^{\infty} D_s & = & \big\{
{\rm all\ connected\ diagrams\ which\ consist\ of\ } N\ge 2 {\rm
\ field} \nonumber \\
& & n{\rm -circles\ of\ bond-coordination\ }\ge 3 
{\rm \ and\ multiple\ }  \nonumber \\
& &{\rm K-{\rm bonds},\ and\ are\ free\ of\ connecting\ circles\  \big\}}
\end{eqnarray}
\end{subequations}
represents the set all remaining completely renormalized graphs.
Under multiple $K$-bonds we mean the possibility of an 
arbitrary number of $K$-bonds between a couple of field circles.
The order of numeration of $D$-diagrams in (24c) is irrelevant, let us say

\noindent
\be \label{25}
\begin{picture}(60,40)(0,7)
    \PhotonArc(20,6)(20,15,165){1}{11}
    \PhotonArc(20,14)(20,195,345){1}{11}
    \Photon(0,10)(40,10){1}{8.5}
    \Vertex(0,10){2} \Vertex(40,10){2}
    \Text(20,-25)[]{$D_1$}
\end{picture}
\begin{picture}(60,40)(0,7)
    \PhotonArc(20,-10)(28,45,135){1}{9}
    \PhotonArc(20,30)(28,225,315){1}{9}
    \PhotonArc(20,6)(20,15,165){1}{11}
    \PhotonArc(20,14)(20,195,345){1}{11}
    \Vertex(0,10){2} \Vertex(40,10){2}
    \Text(20,-25)[]{$D_2$}
\end{picture}
\begin{picture}(60,40)(0,16)
    \PhotonArc(32,6)(32,115,170){1}{7.5}
    \PhotonArc(-12,40)(32,295,355){1}{7.5}
    \PhotonArc(9,5)(32,10,70){1}{7.5}
    \PhotonArc(52,40)(32,185,250){1}{7.5}
    \Photon(0,10)(40,10){1}{8}
    \Vertex(0,10){2} \Vertex(40,10){2} \Vertex(20,35){2}
    \Text(20,-15)[]{$D_3$}
\end{picture}
\SetScale{0.9}
\noindent
\begin{picture}(60,40)(0,13)
    \Photon(0,0)(0,40){1}{7}
    \Photon(40,0)(40,40){1}{7}
    \Vertex(0,0){2} \Vertex(40,0){2} \Vertex(0,40){2} \Vertex(40,40){2}
    \PhotonArc(20,-20)(28,45,135){1}{8}
    \PhotonArc(20,20)(28,225,315){1}{8}
    \PhotonArc(20,20)(28,45,135){1}{8}
    \PhotonArc(20,60)(28,225,315){1}{8}
    \Text(20,-19)[]{$D_4$}
\end{picture} 
\begin{picture}(60,40)(0,12)
    \Photon(0,0)(40,0){1}{7}
    \Photon(0,0)(0,40){1}{7}
    \Photon(0,40)(40,40){1}{7}
    \Photon(40,0)(40,40){1}{7}
    \Photon(0,0)(40,40){1}{8}
    \Photon(0,40)(40,0){1}{8}
    \Vertex(0,0){2} \Vertex(40,0){2} \Vertex(0,40){2} \Vertex(40,40){2}
    \Text(20,-19)[]{$D_5$}
\end{picture}
\SetScale{1}
\ee
\vskip2cm

\noindent and so on.
The symbol $D_s$ will reflect the notation of a given diagram and 
simultaneously its integral representation.

Having classified the renormalized graphs of $\beta \bar F^{ex}$,
we proceed by considering the direct correlation $c$, defined by
equation (5).
From (\ref{24}) one derives
\begin{subequations} \label{26}
\be
c(1,2) = 
\begin{picture}(35,20)(0,7)
    \DashLine(0,10)(35,10){5}
    \BCirc(0,10){2.5} \BCirc(35,10){2.5}
    \Text(0,0)[]{1} \Text(35,0)[]{2}
\end{picture} \ +
c_0(1,2) + \sum_{s=1}^{\infty} c_s(1,2) 
\ee
where $c_0(1,2) = \delta^2 D_0 / \delta n(1) \delta n(2)$ can be easily
shown to correspond to the renormalized ``watermelon'' Meeron graph
\be
c_0(1,2) = \ \ 
\begin{picture}(50,40)(0,7)
    \PhotonArc(20,-10)(28,45,135){1}{9}
    \PhotonArc(20,30)(28,225,315){1}{9}
    \BCirc(0,10){2.5} \BCirc(40,10){2.5}
    \Text(0,0)[]{1} \Text(40,0)[]{2}
\end{picture} = {1\over 2!} K^2(1,2)
\ee
and $c_s(1,2)$ with $s=1,2,\ldots$ denotes the whole family of 
1,2-rooted diagrams generated from $D_s$,
\be
c_s(1,2) = \displaystyle{\delta^2 D_s \over \delta n(1) \delta n(2)} 
\ee
\end{subequations}
To get explicitly a given family $c_s$, one has to take into account 
the functional dependence of the dressed $K$-bonds (\ref{23}) 
on the density as well. 
Since with regard to (23') it holds
\be \label{27}
{\delta K(1,2) \over \delta n(3)} = K(1,3) K(3,2) 
\ee
the functional derivative of $D_s$ with respect to the density field
generates the root circle not only at field-circle positions, but also on 
$K$-bonds, causing their ``correct K-K division''.
For example, in the case of generator $D_1$ drawn in (\ref{25}), we obtain

\noindent
\begin{eqnarray} \label{28} 
c_1(1,2)=\ \
\begin{picture}(55,40)(0,7)
    \PhotonArc(20,6)(20,15,165){1}{11}
    \PhotonArc(20,14)(20,195,345){1}{11}
    \Photon(0,10)(40,10){1}{8.5}
    \BCirc(0,10){2.5} \BCirc(40,10){2.5}
    \Text(-2,-2)[]{1} \Text(42,-2)[]{2}
\end{picture}\ +\ \ \
\begin{picture}(55,40)(0,7)
    \PhotonArc(20,6)(20,15,165){1}{11}
    \PhotonArc(20,14)(20,195,345){1}{11}
    \Photon(0,10)(40,10){1}{8.5}
    \BCirc(0,10){2.5} \BCirc(20,26){2.5}
    \Text(-2,-2)[]{1} \Text(20,36)[]{2} \Vertex(40,10){2.2}
\end{picture}\ +\ \ \
\begin{picture}(55,40)(0,7)
    \PhotonArc(20,6)(20,15,165){1}{11}
    \PhotonArc(20,14)(20,195,345){1}{11}
    \Photon(0,10)(40,10){1}{8.5}
    \BCirc(0,10){2.5} \BCirc(20,26){2.5}
    \Text(-2,-2)[]{2} \Text(20,36)[]{1} \Vertex(40,10){2.2}
\end{picture} \nonumber \\
+\ \ \
\begin{picture}(55,50)(0,7)
    \PhotonArc(20,6)(20,15,165){1}{11}
    \PhotonArc(20,14)(20,195,345){1}{11}
    \Photon(0,10)(40,10){1}{8.5}
    \BCirc(20,26){2.5} \BCirc(20,-6){2.5}
    \Vertex(0,10){2.2} \Vertex(40,10){2.2}
    \Text(20,36)[]{1} \Text(20,-16)[]{2}
\end{picture}\ +\ \ \
\begin{picture}(55,50)(0,7)
    \PhotonArc(20,6)(20,15,165){1}{11}
    \PhotonArc(20,14)(20,195,345){1}{11}
    \Photon(0,10)(40,10){1}{8.5}
    \BCirc(11,24){2.5} \BCirc(29,24){2.5}
    \Vertex(0,10){2.2} \Vertex(40,10){2.2}
    \Text(11,34)[]{1} \Text(29,34)[]{2}
\end{picture} 
\end{eqnarray}
\vskip1cm
\noindent It stands to reason that the coordination of field circles 
remains to be $\ge 3$ after the functional derivation, while the root 
1,2-circles can be two-coordinated (just when being generated on a $K$-bond).
The $\{ c_s \}_{s=1}^{\infty}$ diagram families evidently do not overlap
with each other.

The specialization to the infinite 2d OCP, with dimensionless interaction
energy (\ref{11}), leads to the dressed $K$-bond (\ref{23}) of the form
\begin{subeqnarray} \label{29}
\hat K({\vek k}) & = & - {\Gamma \over k^2+2\pi \Gamma n} \\
K({\vek r}) & = & - \displaystyle{\Gamma \over 2\pi} \int 
\displaystyle{1\over k^2 +2\pi \Gamma n} 
\exp (i{\vek k}\cdot {\vek r}) \rd {\vek k}\hskip3cm \nonumber\\
& = & - \Gamma K_0(r \sqrt{2\pi \Gamma n})
\end{subeqnarray}
The series renormalization of the logarithmic interaction thus
leads to the modified Bessel function of second kind: its decay
to zero at asymptotically large distance makes the renormalized
Mayer diagrams properly convergent$^{(11)}$.
Note the specific $r\sqrt{n}$ dependence of $K$ which has a
fundamental impact on the $n$-classification of renormalized diagrams.
As concerns the $\Gamma$-order of a given diagram $D$ with $N$ field circles
and $L$ bonds, every dressed bond (\ref{29}) brings the factor $\Gamma$ and
enforces the substitution $r'=r\sqrt{\Gamma}$ which manifests itself
as the $\Gamma^{-1}$ factor for each field-circle integration $\sim
\int r dr$, so that the $\Gamma$-order $= L-N$.
For example, in (\ref{25}), $D_1 \sim \Gamma$ and $D_2, D_3, D_4, D_5$ 
constitute the complete set of $\beta \bar F^{ex}$-diagrams $\sim \Gamma^2$.

\vskip 2truecm

\ni {\bf 5. SUM RULES}
\medskip

Let the given completely renormalized diagram $D_s$ $(s=1,\ldots)$ of 
the excess Helmholtz free energy be composed of $N$ skeleton vertices
$i = 1,\ldots, N$ and $L$ bonds $\alpha = 1,\ldots, L$.
$D_s$ can be formally expressed as
\be \label{30}
D_s[n] = \int \prod_{i=1}^N \left[ \rd i \  n(i) \right]
\prod_{\alpha =1}^L K(\alpha_1,\alpha_2)
\ee
where $\alpha_1,\alpha_2 \in \{ 1,\ldots,N \}, \alpha_1 < \alpha_2$
denotes the ordered pair of vertices joint by the $\alpha$-bond.
Whenever not confusing, we will use the symbol $K_{\alpha} \equiv
K(\alpha_1,\alpha_2)$ and omit the ranges $\{ 1,N \}$ and $\{ 1,L \}$
for a sum or product over skeleton vertices $i$ and bonds $\alpha$,
respectively.
The family of direct correlations $c_s({\vek r},{\vek r}')$,
generated according to $c_s({\vek r},{\vek r}') = \delta^2 D_s[n]/
\delta n({\vek r}) \delta n({\vek r}')$, can be straightforwardly
expressed in the uniform-density regime $n(i) =n$ as follows:
$$
\begin{array}{rclr} 
c_s({\vek r},{\vek r}') & = & \displaystyle{n^{N-2} \int \prod_i \rd i
\sum_{i,j \atop (i\ne j)} \delta({\vek r}-i) \delta({\vek r}'-j) \prod_{\alpha}
K_{\alpha}} & (a) \nonumber \\
&  & + \displaystyle{ n^{N-1} \int \prod_i \rd i \sum_i 
\delta({\vek r}-i) \sum_{\alpha} K(\alpha_1,{\vek r}') K({\vek r}',\alpha_2)
\prod_{\beta\ne \alpha} K_{\beta}} & (b) \nonumber \\ 
&  & + \displaystyle{ n^{N-1} \int \prod_i \rd i \sum_i 
\delta({\vek r}'-i) \sum_{\alpha} K(\alpha_1,{\vek r}) K({\vek r},\alpha_2)
\prod_{\beta\ne \alpha} K_{\beta}} & (c) \nonumber \\ 
&  & + \displaystyle{ n^N \int \prod_i \rd i
\sum_{\alpha} K(\alpha_1,{\vek r}')K({\vek r}',{\vek r})
K({\vek r},\alpha_2) \prod_{\beta\ne \alpha} K_{\beta}} & (d) \nonumber \\ 
&  & + \displaystyle{ n^N \int \prod_i \rd i
\sum_{\alpha} K(\alpha_1,{\vek r})K({\vek r},{\vek r}')
K({\vek r}',\alpha_2) \prod_{\beta\ne \alpha} K_{\beta}} & (e) \nonumber \\ 
&  & + \displaystyle{ n^N \int \prod_i \rd i
\sum_{\alpha, \beta \atop (\alpha\ne \beta)} 
K(\alpha_1,{\vek r})K({\vek r},\alpha_2)
K(\beta_1,{\vek r}') K({\vek r}',\beta_2) \prod_{\gamma\ne \alpha,\beta} 
K_{\gamma}} & (f) \nonumber \\ 
\end{array}\eqno(31)
$$
\addtocounter{equation}{1}
\ni where we have applied the functional relation (\ref{27}).
The $(a)$ term on the rhs of (31) corresponds to the creation of
root points at two skeleton vertices, the next two $(b,c)$ terms
to one root circle generated at the skeleton and the other one at a bond,
the $(d,e)$ terms to two root points at the same bond and the last
$(f)$ term represents two root points generated at different 
renormalized bonds $\alpha \ne \beta$.
It stands to reason that now 
$K({\vek r},{\vek r}') = K(\vert {\vek r}-{\vek r}'\vert )$ 
satisfying (A1),(A2a,b) and, consequently, 
$c_s({\vek r},{\vek r}') = c_s(\vert {\vek r}-{\vek r}'\vert )$.

In this section, we aim at proving the validity of the moment equalities
\begin{subequations} \label{32}
\begin{eqnarray}
\int c_s({\vek r}) \rd {\vek r} & = & 0 \\
\int r^2 c_s({\vek r}) \rd {\vek r} & = & 0
\end{eqnarray}
\end{subequations}
for every family $s=1,2, \ldots$, regardless of the topology of
the generating diagram $D_s$.
To keep the interchange-particle symmetry and the 
translational-invariance property of the problem, we will use instead of 
(32a,b) the following equivalent definitions of the moments:
\begin{subequations} \label{33}
\begin{eqnarray}
J_0^{(s)} & = & {1\over V} \int c_s({\vek r},{\vek r}') \rd {\vek r} 
\rd {\vek r}' \\
J_2^{(s)} & = & {1\over V} \int \vert {\vek r} - {\vek r}' \vert^2 
c_s({\vek r},{\vek r}') \rd {\vek r} \rd {\vek r}'
\end{eqnarray}
\end{subequations}
The introduction of volume $V$ into the formulation of the infinite-volume
limit does not mean any loss of rigour.
Definitions (33a,b) have to be understood in the sense that an arbitrary
one of integration coordinates ${\vek r}, {\vek r}', \{ i\}$ can be taken,
due to the invariance of the integrated function with respect to a
uniform shift in all variables, as a reference and put at the origin 
${\vek 0}$, with the simultaneous cancellation of $V$. 
The right choice of the reference can simplify otherwise tedious
algebra.

\vskip 1truecm

\ni {\bf 5.1. Proof of the zeroth-moment condition (32a)}
\medskip

By using the definition (33a), the zeroth moment of $c_s$ (31) 
can be expressed as
$$
\begin{array}{rclr}
J_0^{(s)} & = & \displaystyle{ N (N-1) n^{N-2} {1\over V}
\int \prod_i \rd i \prod_{\alpha} K_{\alpha}} & (a) \nonumber
\\ & & \displaystyle{
+ 2 N n^{N-1} {1\over V} \int \prod_i \rd i \sum_{\alpha}
{\partial K_{\alpha} \over \partial n} \prod_{\beta\ne \alpha}
K_{\beta}} & (b+c) \nonumber
\\ & & \displaystyle{
+ n^N {1\over V} \int \rd {\vek r} \int \prod_i \rd i 
\sum_{\alpha} {\partial K(\alpha_1,{\vek r})\over \partial n}
K({\vek r},\alpha_2) \prod_{\beta\ne \alpha}K_{\beta}} & (d) \nonumber 
\\ & & \displaystyle{
+ n^N {1\over V} \int \rd {\vek r} \int \prod_i \rd i 
\sum_{\alpha} K(\alpha_1,{\vek r}) {\partial K({\vek r},\alpha_2)
\over \partial n} \prod_{\beta\ne \alpha}K_{\beta}} & (e) \nonumber
\\ & & \displaystyle{
+ n^N {1\over V} \int \prod_i \rd i \sum_{\alpha, \beta \atop
(\alpha\ne \beta)} {\partial K_{\alpha} \over \partial n}  
{\partial K_{\beta} \over \partial n}  
\prod_{\gamma\ne \alpha,\beta} K_{\gamma}} & (f) \nonumber \\
\end{array}\eqno(34)
$$
\addtocounter{equation}{1}
\ni where we have taken into account relation (A2a).
Since
$$
\int \rd {\vek r} \left[ {\partial K(\alpha_1,{\vek r})\over \partial n}
K({\vek r},\alpha_2) + K(\alpha_1,{\vek r}) {\partial K({\vek r},\alpha_2)
\over \partial n} \right] = {\partial^2 K(\alpha_1,\alpha_2) \over
\partial n^2}$$
\ni we find
\be \label{35}
J_0^{(s)} = {\partial^2 \over \partial n^2} \left[
{n^N \over V} \int \prod_{i=1}^N \rd i \prod_{\alpha=1}^L K_{\alpha}
\right] 
\ee
Let us put say $i=1$ at the origin ${\vek 0}$, and ``cancel'' the
integration over 1 with volume $V$.
As $K(\alpha_1,\alpha_2) = - \Gamma K_0(\sqrt{2\pi\Gamma n} \vert
\alpha_1-\alpha_2 \vert)$ for the 2d OCP, the evoked substitution
$i' = i \sqrt{2\pi\Gamma n}$ for $(N-1)$ remaining coordinates 
$i=2,\ldots,N$ results in the factor $1/n^{(N-1)}$.
Therefore,
\be \label{36}
J_0^{(s)} \sim {\partial^2 \over \partial n^2} 
{n^N \over n^{(N-1)}} = 0
\ee
in agreement with (32a).
\newpage	
\ni {\bf 5.2. Proof of the second-moment condition (32b)}
\medskip

By using the definition (33b), the second moment of $c_s$ (31) 
is written as
$$
\begin{array}{rclr}
J_2^{(s)} & = & \displaystyle{
{n^{N-2} \over V} \int \prod_i \rd i \sum_{i,j\atop (i\ne j)}
\vert i-j \vert^2 \prod_{\alpha} K_{\alpha}
} & (a) \nonumber \\
& & \displaystyle{
+ {2 n^{N-1} \over V} \int \prod_i \rd i \sum_{i,\alpha} \int 
\vert {\vek r} - i \vert^2 K(\alpha_1,{\vek r})
K({\vek r},\alpha_2) \rd {\vek r} \prod_{\beta\ne \alpha} K_{\beta}
} & (b+c) \nonumber \\
& & \displaystyle{
+ {2 n^N \over V} \int \prod_i \rd i \sum_{\alpha} \int \vert {\vek r}
- {\vek r}' \vert^2  K(\alpha_1,{\vek r}) K({\vek r},{\vek r}')
K({\vek r}',\alpha_2) \rd {\vek r} \rd {\vek r}'
\prod_{\beta\ne \alpha} K_{\beta}
} & (d+e) \nonumber \\
& & \displaystyle{
+ {n^N \over V} \int \prod_i \rd i \sum_{\alpha,\beta \atop (\alpha
\ne \beta)} \int  \vert {\vek r} - {\vek r}' \vert^2  K(\alpha_1,{\vek r}) 
K({\vek r},\alpha_2)} \nonumber \\
& & \hskip 6cm
\times K(\beta_1,{\vek r}') K({\vek r}',\beta_2) 
\rd {\vek r} \rd {\vek r}' \prod_{\gamma\ne \alpha,\beta} K_{\gamma}
 & (f) \nonumber \\
\end{array}\eqno(37)
$$
\addtocounter{equation}{1}
\ni The integrations over {\vek r} and {\vek r}' in (37) correspond
to root points generated on renormalized $K$-bonds, and interacting
with another root point at the skeleton or with one another.
In Appendix A we show how to transform these integrals to the form
with exclusively skeleton $\vert i-j \vert^2, \Phi(\vert i-j \vert)$
and $\Psi(\vert i-j \vert)$ interactions [for definitions of $\Phi$
and $\Psi$ see (A7) and (A10), respectively] and appropriately
``decorated'' bonds [under decoration, we mean the derivation with
respect to density, equation (A2a)].
The successive application of formulae (A8), (A10) and (A9) to the
respective terms (b+c), (d+e) and (f) of (37) yields
$$
\begin{array}{rclr}
J_2^{(s)} & = & \displaystyle{
{n^{N-2} \over V} \int \prod_i \rd i \sum_{i,j\atop (i\ne j)}
\vert i-j \vert^2 \prod_{\alpha} K_{\alpha}
} & (a) \nonumber \\
& & \displaystyle{
+ {2 n^{N-1} \over V} \int \prod_i \rd i \sum_{\alpha} 
\left[ \sum_i {1\over 2} \left( \vert \alpha_1 -i \vert^2 + \vert
\alpha_2 -i \vert^2 \right) {\partial K_{\alpha} \over \partial n}
+ N \Phi_{\alpha} \right] \prod_{\beta\ne \alpha} K_{\beta}
} & (b+c) \nonumber \\
& & \displaystyle{
+ {2 n^N \over V} \int \prod_i \rd i \sum_{\alpha} 
\Psi_{\alpha} \prod_{\beta\ne \alpha} K_{\beta}
} & (d+e) \nonumber \\
& & + \displaystyle{{n^N \over V}} \int \prod_i \rd i 
\sum_{\alpha,\beta \atop (\alpha\ne \beta)} 
\big[ \displaystyle{{1\over 4}} \left( 
\vert \alpha_1 -\beta_1 \vert^2 +\vert \alpha_1 -\beta_2 \vert^2 +
\vert \alpha_2 -\beta_1 \vert^2 +\vert \alpha_2 -\beta_2 \vert^2 
\right) \nonumber \\ & & \hskip 4cm \times \displaystyle{
{\partial K_{\alpha} \over \partial n}{\partial K_{\beta} \over \partial n} 
+ {\partial K_{\alpha} \over \partial n} \Phi_{\beta}
+ {\partial K_{\beta} \over \partial n} \Phi_{\alpha}} \big]
\displaystyle{\prod_{\gamma\ne \alpha,\beta} K_{\gamma}}
& (f) \nonumber \\
\end{array}\eqno(38)
$$
\addtocounter{equation}{1}
\quad Let us now define the auxiliary function
\begin{eqnarray} \label{39}
G^{(s)}(n) & = & {2n^{N+1}\over V} \int \prod_i \rd i \sum_{\alpha}
\Phi_{\alpha} \prod_{\beta\ne \alpha} K_{\beta} 
\nonumber \\ & \equiv &
{n^{N+1}\over V} \int \prod_i \rd i \left( \sum_{\alpha}
\Phi_{\alpha} \prod_{\beta\ne \alpha} K_{\beta} 
+ \sum_{\beta} \Phi_{\beta} \prod_{\alpha\ne \beta} K_{\alpha} 
\right) 
\end{eqnarray}
Up to the prefactor $2/V$, it originates from $D_s(n)$ (30)
by picking out successively bonds one after the other and 
interchanging the bond factor $K \to n \Phi$.  
$\Phi_{\alpha}$ is expressible from (A7) and (29a) in the form
\be \label{40}
\Phi_{\alpha} = {4\Gamma^2 \over (2\pi \Gamma n)^2}
\int {p^2 \over (p^2 +1)^4} \exp{ \left[ i {\vek p}\cdot
(\alpha_1 -\alpha_2) \sqrt{2\pi \Gamma n} \right]}
\rd {\vek p}
\ee
With the aid of the scaling argument leading to relation (36),
$G^{(s)}$ can be shown to scale with $n$ like
\be \label{41}
G^{(s)}(n) \sim {n^{N+1} \over n^{N-1} n^2} = n^0
\ee
As a consequence, $\partial G^{(s)}(n) /\partial n = 0$.
Explicitly,
\begin{eqnarray} \label{42}
0 & = & {2(N+1)n^N \over V}
\int \prod_i \rd i \sum_{\alpha} \Phi_{\alpha} 
\prod_{\beta\ne \alpha} K_{\beta} \nonumber \\
& & + {2n^{N+1} \over V} \int \prod_i \rd i \sum_{\alpha}
{\partial \Phi_{\alpha} \over \partial n} \prod_{\beta\ne\alpha}
K_{\beta} \nonumber \\
& & + {n^{N+1}\over V} \int \prod_i \rd i \sum_{\alpha,\beta \atop (\alpha
\ne \beta)} \left( \Phi_{\alpha} {\partial K_{\beta} \over \partial n}+
\Phi_{\beta} {\partial K_{\alpha}\over \partial n} \right)
\prod_{\gamma\ne\alpha,\beta} K_{\gamma}
\end{eqnarray}
Subtracting $\{ {\rm Eq.(42)} \}/n$ from formula (38), 
the latter takes a simpler form
\begin{eqnarray} \label{43}
J_2^{(s)} & = & 
{n^{N-2} \over V} \int \prod_i \rd i \sum_{i,j\atop (i\ne j)}
\vert i-j \vert^2 \prod_{\alpha} K_{\alpha}
\nonumber \\ & & 
+ {n^{N-1} \over V} \int \prod_i \rd i \sum_{i,\alpha} 
\left( \vert \alpha_1 -i \vert^2 + \vert \alpha_2 -i \vert^2 \right) 
{\partial K_{\alpha} \over \partial n} \prod_{\beta\ne \alpha} K_{\beta}
\nonumber \\ & & 
+ {2 n^N \over V} \int \prod_i \rd i \sum_{\alpha} 
\left[ \Psi_{\alpha}- {1\over n} 
{\partial( n \Phi_{\alpha})\over \partial n} \right]
\prod_{\beta\ne \alpha} K_{\beta}
\nonumber \\ & & 
+ {n^N \over 4 V} \int \prod_i \rd i 
\sum_{\alpha,\beta \atop (\alpha\ne \beta)} 
\left( 
\vert \alpha_1 -\beta_1 \vert^2 +\vert \alpha_1 -\beta_2 \vert^2 +
\vert \alpha_2 -\beta_1 \vert^2 +\vert \alpha_2 -\beta_2 \vert^2 
\right) \nonumber \\ & & 
\hskip 4cm \times 
{\partial K_{\alpha} \over \partial n}{\partial K_{\beta} \over \partial n} 
\prod_{\gamma\ne \alpha,\beta} K_{\gamma}
\end{eqnarray}
It is trivial to show that
$$
\begin{array}{rcl}
\Psi_{\alpha} - \displaystyle{{1\over n} {\partial (n \Phi_{\alpha}) 
\over \partial n}}
& = &  \displaystyle{ - \int \e^{i{\vekexp p}\cdot (\alpha_1-\alpha_2)}
\Delta_{{\vekexp p}}  
\left[ {\Gamma^2 \over 4 n (p^2+2\pi\Gamma n)^2}
-{\pi\Gamma^3 \over (p^2+2\pi\Gamma n)^3} \right] \rd {\vek p}} \\
& = & \displaystyle{ \vert \alpha_1 - \alpha_2 \vert^2 
\int \e^{i{\vekexp p}\cdot (\alpha_1-\alpha_2)}
\left[ {\Gamma^2 \over 4 n (p^2+2\pi\Gamma n)^2}
-{\pi\Gamma^3 \over (p^2+2\pi\Gamma n)^3} \right] \rd {\vek p}} \\
& = & \displaystyle{{\vert \alpha_1 - \alpha_2 \vert^2 \over 4 n^2}
\left( n {\partial \over \partial n} \right)^2 K_{\alpha}}
\end{array}\eqno(43')
$$
\ni 
Equation (43) together with the complementary one (43') represent
the formulation of the second moment of $\hat c_s$ in the renormalized 
format in terms of quadratic interactions exclusively
between pairs of skeleton vertices.

For tactical reasons, we now specify the connectivity of diagram
$D_s$, instead of enumerating present renormalized bonds $\alpha
=1,\ldots, L$, by the set $\{ \nu_{ij} \}_{i,j=1 \atop (i\ne j)}^N$ 
where $\nu_{ij} = \nu_{ji}$ is the number of $K$-bonds between skeleton 
vertices $i,j$ ($\nu_{ij} =0$ if there is no bond between $i,j$).
Let us choose a couple of skeleton vertices, say 1 and 2, and
group all factors in (43), (43') associated with the $\vert 1-2 \vert^2$
interaction:
\begin{eqnarray}\label{44}
& & {n^{N-2} \over V} \int \prod_i \rd i ~ \vert 1-2 \vert^2
\prod_{u,v=3\atop (u<v)}^N K^{\nu_{uv}}(u,v) \nonumber \\
& \times & \Big\{ 2 K^{\nu_{12}}(1,2) 
\prod_{i=3}^N K^{\nu_{1i}}(1,i)
\prod_{j=3}^N K^{\nu_{2j}}(2,j) \nonumber \\
& + & 2 \nu_{12} K^{\nu_{12}-1}(1,2) \left[ n {\partial K(1,2)\over 
\partial n} \right] \prod_{i=3}^N K^{\nu_{1i}}(1,i)
\prod_{j=3}^N K^{\nu_{2j}}(2,j) \nonumber \\
& + & K^{\nu_{12}}(1,2) \left[ n {\partial \over \partial n}
\prod_{i=3}^N K^{\nu_{1i}}(1,i)\right]
\prod_{j=3}^N K^{\nu_{2j}}(2,j) \nonumber \\
& + & K^{\nu_{12}}(1,2)
\prod_{i=3}^N K^{\nu_{1i}}(1,i) \left[ n {\partial \over \partial n}
\prod_{j=3}^N K^{\nu_{2j}}(2,j) \right] \nonumber \\
& + & {1\over 2} \nu_{12} K^{\nu_{12}-1}(1,2)\left[ \left(
n {\partial \over \partial n} \right)^2 K(1,2) \right]
\prod_{i=3}^N K^{\nu_{1i}}(1,i)
\prod_{j=3}^N K^{\nu_{2j}}(2,j) \nonumber \\
& + & {1\over 2} \nu_{12} (\nu_{12}-1) K^{\nu_{12}-2}(1,2)
\left[ n {\partial K(1,2) \over \partial n} \right]^2
\prod_{i=3}^N K^{\nu_{1i}}(1,i)
\prod_{j=3}^N K^{\nu_{2j}}(2,j) \nonumber \\
& + & {1\over 2} \nu_{12} K^{\nu_{12}-1}(1,2) \left[ n {\partial
K(1,2)\over \partial n} \right]
\left[ n {\partial \over \partial n}\prod_{i=3}^N K^{\nu_{1i}}(1,i)\right]
\prod_{j=3}^N K^{\nu_{2j}}(2,j) \nonumber \\
& + & {1\over 2} \nu_{12} K^{\nu_{12}-1}(1,2) \left[ n {\partial
K(1,2)\over \partial n} \right]
\prod_{i=3}^N K^{\nu_{1i}}(1,i)
\left[ n {\partial \over \partial n} \prod_{j=3}^N K^{\nu_{2j}}(2,j) 
\right] \nonumber \\
& + & {1\over 2} K^{\nu_{12}}(1,2)
\left[ n {\partial \over \partial n} \prod_{i=3}^N K^{\nu_{1i}}(1,i)\right]
\left[ n {\partial \over \partial n} \prod_{j=3}^N K^{\nu_{2j}}(2,j) 
\right] \Big\} 
\end{eqnarray}
\ni In the case of the considered Bessel functions (\ref{29}), the operator
$n(\partial/\partial n)$ acting on $K(i,j)$ can be substituted by
a coordinate-operator as follows
\begin{eqnarray}\label{45}
n{\partial \over \partial n} K(i,j) & = &
{1\over 2} \left( r_{ij} {\partial \over \partial r_{ij}} \right)
K(\vert {\vek r}_i - {\vek r}_j \vert) \nonumber \\
& \equiv & {1\over 2} \bR_{ij} 
K(\vert {\vek r}_i - {\vek r}_j \vert) 
\end{eqnarray}
\ni As one can derive directly from the definition (\ref{45}), there exist
more equivalent representations of operator $\bR_{ij}$,
\begin{eqnarray}\label{46}
\bR_{ij} & = & r_{ij} {\partial \over \partial r_{ij}} \nonumber \\
& = & ({\vek r}_i - {\vek r}_j)\cdot \nabla_i \nonumber \\
& = & ({\vek r}_j - {\vek r}_i)\cdot \nabla_j \nonumber \\
& = & {1\over 2} ({\vek r}_i - {\vek r}_j)\cdot (\nabla_i 
- \nabla_j) 
\end{eqnarray}
Finally, denoting $F_{ij} = K^{\alpha_{ij}}(\vert i-j\vert)$ and
summing over all pairs of skeleton vertices, Eqs. (43), (44) and (45)
imply
\begin{eqnarray}\label{47}
J_2^{(s)}/n^{N-2} & = & {1\over V} \int \prod_i \rd i
\prod_{u<v} F_{uv} \sum_{i<j} \vert i-j \vert^2 \nonumber \\
& \times & \Big\{ 2 + {1\over F_{ij}} \left( \bR_{ij} F_{ij} \right)
+ {1\over 8 F_{ij}} \left( \bR^2_{ij} F_{ij} \right) \nonumber \\
& + & {1\over 2} \sum_{k\ne i,j} \left[ {1\over F_{ik}} \left( 
\bR_{ik}F_{ik} \right) + {1\over F_{jk}} \left( \bR_{jk}F_{jk} \right)
\right] \nonumber \\
& + & {1\over 8 F_{ij}} \left( \bR_{ij} F_{ij} \right)
\sum_{k\ne i,j} \left[ {1\over F_{ik}} \left( 
\bR_{ik}F_{ik} \right) + {1\over F_{jk}} \left( \bR_{jk}F_{jk} \right)
\right] \nonumber \\
& + & {1\over 8} \sum_{k,l\ne i,j} {1\over F_{ik} F_{jl}}
\left( \bR_{ik} F_{ik} \right)
\left( \bR_{jl} F_{jl} \right) \Big\}
\end{eqnarray}

In what follows we aim at proving the nullity of the rhs of (\ref{47}),
{\it irrespective of} the particular form of the bond-dependent functions
$F_{ij}(r_{ij})$ (provided that the integrals exist what certainly
applies to our case of $F$-functions).
As shown in Appendix B, due to a scaling transformation of coordinates
in integrals of translationally-invariant functions over infinite
$2d$ space, Eq.(\ref{47}) is reducible to a simpler relation
\begin{eqnarray}\label{48}
8 J_2^{(s)}/n^{N-2}  & = &  \int \prod_i \rd i
\prod_{u<v} F_{uv} \sum_{i<j} \vert i-j \vert^2 \nonumber \\
& \times & \Big\{ \sum_{k\ne i,j} \left[
{1\over 2F_{ik}} ({\vek r}_i - {\vek r}_j) \cdot \nabla_i F_{ik}+
{1\over 2F_{jk}} ({\vek r}_j - {\vek r}_i) \cdot \nabla_j F_{jk}
\right] \nonumber \\
& + & \sum_{k,l\ne i,j} {1\over F_{ik} F_{jl}}
\left[({\vek r}_j-{\vek r}_k)\cdot \nabla_i F_{ik} \right]
\left[({\vek r}_i-{\vek r}_l)\cdot \nabla_j F_{jl} \right] \Big\} 
\end{eqnarray}
\ni Let us consider $\{ F_{ij} \}$ to be the functions of $r_{ij}^2$
rather than $r_{ij}$, without changing the symbol $F$.
Thus
\be \label{49}
\nabla_i F_{ij}(r_{ij}^2) = {\rd F_{ij}(r_{ij}^2) \over \rd (r_{ij}^2)} 
2 ({\vek r}_i - {\vek r}_j)
\ee
Using the notation
\be \label{50}
\bF_{ij} = {1\over F_{ij}} {\rd F_{ij}(r_{ij}^2) \over \rd (r_{ij}^2)} 
\ee
Eq.(\ref{49}) can be rewritten as follows
$$\nabla_i F_{ij} = 2 ({\vek r}_i - {\vek r}_j) F_{ij} \bF_{ij}
\eqno(49')$$
Inserting (49') into (\ref{48}) and grouping the $\bF$ and 
$\bF \bF$ terms we obtain
\begin{eqnarray}\label{51}
8 J_2^{(s)}/n^{N-2} & = & \int \prod_i \rd i \prod_{u<v} F_{uv} \big\{
\sum_{(i<j)} \bF_{ij} \sum_{k\ne i,j} {\vek r}_{ij} \cdot
(r_{ik}^2 {\vek r}_{ik} - r_{jk}^2 {\vek r}_{jk}) \nonumber \\
& + & 2 \sum_{(i<j)} \sum_{(k<l)} \bF_{ij} \bF_{kl} \left[ 
r_{ik}^2 ({\vek r}_{ij} \cdot {\vek r}_{kj})
({\vek r}_{il} \cdot {\vek r}_{kl})+
r_{il}^2 ({\vek r}_{ij} \cdot {\vek r}_{lj})
({\vek r}_{ik} \cdot {\vek r}_{lk}) \right. \nonumber \\
&  &\quad \quad \quad \quad \quad \quad 
\left. + r_{jk}^2 ({\vek r}_{ji} \cdot {\vek r}_{ki})
({\vek r}_{jl} \cdot {\vek r}_{kl})+
r_{jl}^2 ({\vek r}_{ji} \cdot {\vek r}_{li})
({\vek r}_{jk} \cdot {\vek r}_{lk}) \right] \big\}
\end{eqnarray}
with the obvious notation ${\vek r}_{ij} = {\vek r}_i - {\vek r}_j$.

Our further goal is to prove the nullity of the rhs of (\ref{51}).
We show that the per-partes integration of the bilinear $\bF \bF$
term in (\ref{51}) gives a linear $\bF$ term which exactly eliminates
the linear $\bF$ term in (\ref{51}).
To do so, we express the bilinear $\bF \bF$ term in 
integral (\ref{51}) in an ``ansatz'' form
\begin{eqnarray} \label{52}
& & \int \prod_i \rd i \prod_{u<v} F_{uv} \left[
({\vek r}_{12} \bF_{12} + {\vek r}_{13} \bF_{13} +
\ldots + {\vek r}_{1N} \bF_{1N} )\cdot \vQ_1 \right. \nonumber \\
& & \left. \quad \quad \quad \quad \quad \quad
+ ({\vek r}_{21} \bF_{12} + {\vek r}_{23} \bF_{23} +
\ldots + {\vek r}_{2N} \bF_{2N} )\cdot \vQ_2 + \ldots \right]
\end{eqnarray}
The vector functions $\vQ_i$ must be linear combinations of $\bF_{kl}$
with $k,l\ne i$ since the terms $\bF_{ij} \bF_{ij}$ vanish in (\ref{51});
$\vQ_1 = \vQ_1^{(2,3)} \bF_{23} + \vQ_1^{(2,4)} \bF_{24} + \ldots
+ \vQ_1^{(N-1,N)} \bF_{N-1,N}$, etc.
In general,
\be \label{53}
\vQ_i = \sum_{k,l\ne i \atop k<l} \vQ_i^{(k,l)} \bF_{kl}
\ee
where the three-point vector coefficients $\vQ_i^{(k,l)}$,
\begin{subequations} \label{54}
\be
\vQ_i^{(k,l)} \ne 0 \quad \quad {\rm iff}\ i\ne k\ne l
\ee
depend only on coordinates $\{ {\vek r}_i, {\vek r}_k, {\vek r}_l \}$.
It is natural to extend their definition as follows
\be
\vQ_i^{(k,l)} = \vQ_i^{(l,k)}
\ee
\end{subequations}
Comparing Eq.(52) with (51), the coefficients to the $\bF_{ik} \bF_{il}$
$(i\ne k\ne l)$ and $\bF_{ij} \bF_{kl}$ $(i\ne j\ne k\ne l)$ terms imply
the respective restrictions on $\vQ$-vectors:
\begin{subequations} \label{55}
\be
{\vek r}_{ki} \cdot \vQ_k^{(i,l)} + {\vek r}_{li}\cdot \vQ_l^{(i,k)}
= 4 r_{kl}^2 ({\vek r}_{ki}\cdot {\vek r}_{li})^2
\ee
\begin{eqnarray}
& & 
{\vek r}_{ij} \cdot \vQ_i^{(k,l)}+{\vek r}_{ji} \cdot \vQ_j^{(k,l)}+
{\vek r}_{kl} \cdot \vQ_k^{(i,j)}+{\vek r}_{lk} \cdot \vQ_l^{(i,j)}
\nonumber \\
& &  \quad \quad \quad
= 4 \left[ r_{ik}^2 ({\vek r}_{ij} \cdot {\vek r}_{kj})
({\vek r}_{il} \cdot {\vek r}_{kl})+
r_{il}^2 ({\vek r}_{ij} \cdot {\vek r}_{lj})
({\vek r}_{ik} \cdot {\vek r}_{lk}) \right. \nonumber \\
&  &\left.\quad \quad \quad \quad 
+ r_{jk}^2 ({\vek r}_{ji} \cdot {\vek r}_{ki})
({\vek r}_{jl} \cdot {\vek r}_{kl})+
r_{jl}^2 ({\vek r}_{ji} \cdot {\vek r}_{li})
({\vek r}_{jk} \cdot {\vek r}_{lk}) \right] 
\end{eqnarray}
\end{subequations}
\ni with no order inequalities put on $\{ i,j,k,l \}$.
The important feature of the ansatz proposal (\ref{52}) consists in
the equality
\be \label{56}
\prod_{u<v} F_{uv} \left( 
{\vek r}_{i,1} \bF_{i1} + \ldots +{\vek r}_{i,N} \bF_{iN} \right) 
= {1\over 2} \nabla_i \left( \prod_{u<v} F_{uv} \right)
\ee
[see relation (49')].
Using formula (\ref{56}) for every term in (\ref{52}), the
consequent per-partes integrations lead to
\be \label{57}
(\ref{52}) = - {1\over 2} \int \prod_i \rd i \prod_{u<v} F_{uv}
\sum_i \nabla_i \cdot \vQ_i
\ee
The point is that the functions $\{ \bF_{kl} \}$ in $\vQ_i$ (\ref{53})
do not depend on coordinate ${\vek r}_i$, and therefore the nabla
operator acts only on coefficients $\{ \vQ_i^{(k,l)} \}$.
This is why expression (\ref{52}) becomes linear in $\bF$-functions.
It compensates the linear term in (\ref{51}) just when
\be \label{58}
\nabla_i \cdot \vQ_i^{(k,l)} = 2 {\vek r}_{kl} \cdot
(r_{ki}^2 {\vek r}_{ki} - r_{li}^2 {\vek r}_{li})
\ee
To summarize, $J_2^{(s)}$, given by (\ref{51}), equals to zero provided 
there exists a three-point vector function $\vQ_i^{(k,l)}$ with
properties (54a), (54b) and satisfying conditions (55a), (55b)
and (\ref{58}).
One can readily verify on computer that such vector function
exists: it is the homogeneous polynomial of the fifth order and
its $x$-component reads
\begin{eqnarray} \label{59}
\left[ \vQ_i^{(k,l)} \right]_x & = & {2\over 3} \left( 
- 2 u_x^4 v_x - 2 u_x v_x^4 + 2 u_x^3 v_x^2 + 2 u_x^2 v_x^3
- 2 u_x^3 v_y^2 - 2 u_y^2 v_x^3 - u_x v_y^4 - u_y^4 v_x \right. \nonumber \\
& & \quad \left.
-u_x^3 u_y v_y - u_y v_x^3 v_y - 3 u_x v_x^2 v_y^2 - 3 u_x^2 u_y^2 v_x
- u_x u_y^3 v_y - u_y v_x v_y^3 \right. \nonumber \\
& & \quad \left.
+ 6 u_x u_y^2 v_x^2 + 6 u_x^2 v_x v_y^2 - 6 u_x u_y^2 v_y^2
-6 u_y^2 v_x v_y^2 + 8 u_x u_y v_y^3 + 8 u_y^3 v_x v_y 
\right)
\end{eqnarray} 
where ${\vek u} = {\vek r}_k - {\vek r}_i$ and ${\vek v} = {\vek r}_l
- {\vek r}_i$.
The $y$-component $[\vQ_i^{(k,l)}]_y$ results from $(\ref{59})$
under interchange transformation
$u_x \leftrightarrow u_y, v_x \leftrightarrow v_y$.
We conclude that $J_2^{(s)} = 0$, confirming relation (32b).
\vskip 2truecm

\ni {\bf 6. CONCLUSION}
\medskip

The Fourier transform of (26a) results in
\be \label{60}
\hat c({\vek k}) = - \Gamma /k^2 + \hat c_0({\vek k}) +
\sum_{s=1}^{\infty} \hat c_s({\vek k})
\ee 
With regard to the equalities (32a,b) proved in the previous section,
we have
\be \label{61}
\hat c_s({\vek k}) = O(k^4) \quad \quad \quad (s=1,2,\ldots)
\ee
Using the formula$^{(25)}$
\be \label{62}
\int_0^{\infty} x^{1+2s} K_0^2(x) \rd x = 2^{(2s-1)}
{(s!)^4 \over (2s+1)!} \quad \quad \quad (s\ge 0)
\ee
the contribution of the renormalized Meeron-graph (26b) to the Fourier
component of the direct correlation reads
\be \label{63}
\hat c_0({\vek k}) = {\Gamma \over 8 \pi n} - {k^2 \over 96 (\pi n)^2}
+ O(k^4)
\ee
Consequently, the expansion of $\hat c({\vek k})$ up to the $k^2$-term
coincides with the suggested formula (\ref{21}) and, via the OZ relation,
the (rescaled) sixth moment of $h$ is indeed a finite $\Gamma$-truncation
\be \label{64}
n \left( {\pi \Gamma n \over 2} \right)^3 \int r^6 h({\vek r}) \rd {\vek r}
= {3\over 4} (\Gamma -6) (8-3\Gamma) 
\ee
as indicated in (\ref{20}).

In conclusion, we would like to stress that the derivation of the sixth
moment of $h$ was possible due to the property (\ref{61}), valid separately
for {\it each} family of direct-correlation diagrams generated from 
the corresponding, completely renormalized, graph of the Helmholtz
free energy.
The higher-order coefficients of the ${\vek k}$-expansion of
$\hat c_s({\vek k})$ $(s=1,2, \ldots)$ we were able to attain do not
longer vanish, e.g.,
\be \label{65}
\hat c_1({\vek k}) = 
- {k^4 \over (2\pi n)^3} {2\over (4!)^2 5!} \int_0^{\infty}
K_0^3(x) [176 x + 108 x^3 + 9 x^5] \rd x + O(k^6)
\ee
where the numerical value of the integral $\approx 116.68\ldots$.
This fact indicates that the eight- and higher-order (appropriately rescaled) 
moments of $h$ are probably infinite series in $\Gamma$.
However, there might exist another mechanism for the exact solvability
of higher-order moments as well.
We believe that the present method will answer this interesting
question. 

\vskip 1truecm


\ni {\bf Acknowledgment}
\medskip

\ni We are grateful to Jerry Percus for stimulating discussions
about the subject.
This work was supported by Grant VEGA No. 2/4109/97.

\vskip 2truecm

\ni {\bf APPENDIX A}
\medskip

Here, we derive an algebraic procedure removing a two-coordinated
root point, generated by the functional derivation with respect to
density at a renormalized $K$-bond and interacting quadratically with 
another root point (generated either at a skeleton more-than-two-coordinated
vertex or at a bond), in the representation of the second moment of 
a $c_s$-family (subsection 5.2).
We consider the uniform regime with constant density $n$ and
translationally+rotationally invariant interactions
$$K({\vek r}_1,{\vek r}_2) = \int \exp[i {\vek p}\cdot ({\vek r}_1 - 
{\vek r}_2)] \hat K({\vek p}) {\rd {\vek p} \over 2\pi}
\eqno(A1)$$
$K$ as the function of $n$ is given by the uniform analogue of
functional relation (\ref{27}),
$${\partial K(\vert {\vek r}_1 - {\vek r}_2 \vert) \over \partial n}
= \int K(\vert {\vek r}_1 - {\vek r} \vert) K(\vert {\vek r} - 
{\vek r}_2 \vert) \rd {\vek r} \eqno(A2a)$$
or, in the 2d Fourier picture,
$${\partial \hat K({\vek p}) \over \partial n} = 2 \pi
\hat K^2({\vek p}) \eqno(A2b)$$

Let us first consider the case represented schematically as follows
$$
f({\vek u}_1,{\vek u}_2) =
\begin{picture}(70,50)(-4,7)
    \Photon(10,10)(50,10){1}{8.5}
    \Photon(10,10)(0,10){1}{2}
    \Photon(10,10)(2,18){1}{2}
    \Photon(10,10)(2,2){1}{2}
    \Photon(50,10)(60,10){1}{2}
    \Photon(50,10)(58,18){1}{2}
    \Photon(50,10)(58,2){1}{2}
    \BCirc(10,10){2.5}
    \BCirc(50,10){2.5}
    \DashLine(30,10)(30,-20){1}
    \Photon(30,-20)(30,-30){1}{2}
    \Photon(30,-20)(22,-28){1}{2}
    \Photon(30,-20)(38,-28){1}{2}
    \BCirc(30,-20){2.5}
    \Vertex(30,10){2.2}
    \Text(14,21)[]{$\vek u_1$}
    \Text(48,21)[]{$\vek u_2$}
    \Text(30,22)[]{$\vek r$}
    \Text(37,-16)[]{$\vek 0$}
\end{picture} 
= \int r^2 K(\vert {\vek u}_1 - {\vek r} \vert)
K(\vert {\vek r} - {\vek u}_2 \vert) \rd {\vek r} \eqno(A3)$$

\bigskip

\bigskip
\noindent
where ${\vek u}_1$ and ${\vek u}_2$ skeleton vectors define the bond
decorated via (\ref{27}) by the two-coordinated root point ${\vek r}$, 
integrated out and interacting via $r^2$-interaction with another root point 
(whose coordination is irrelevant at this stage), put for simplicity 
at the origin {\vek 0}.
Using the Fourier representation of $K$, (A3) can be rewritten in the form
$$
\begin{array}{lcl}
f({\vek u}_1,{\vek u}_2) & = & \displaystyle{-\frac{1}{2}\int \rd {\vek r} 
\int \frac{\rd \vek p}{2\pi} \int \frac{\rd \vek q}{2\pi}}
~\hat K({\vek p})~\hat K({\vek q}) \\ & &
\times
\big\{ \e^{i{\vekexp p}\cdot{\vekexp u}_1}
\e^{i {\vekexp q}\cdot({\vekexp r}-{\vekexp u}_2)}
\Delta_p [\e^{-i{\vekexp p}\cdot{\vekexp r}}] \nonumber 
+ \e^{i{\vekexp p}\cdot({\vekexp u}_1-{\vekexp r})}
\e^{-i {\vekexp q}\cdot{\vekexp u}_2}
\Delta_q [\e^{i{\vekexp q}\cdot{\vekexp r}}] \big\} \\ 
\end{array} \eqno(A4)
$$
Integrating twice per partes in (A4) over $p$ $(q)$ [what can be
certainly done for our $\hat K$ (29a) with vanishing boundary
contributions] and then integrating over ${\vek r}$, implying
$\delta({\vek p} - {\vek q})$, we get
$$
\begin{array}{rcl}
f({\vek u}_1,{\vek u}_2) & = & - \displaystyle{\int} \rd {\vek p} \left\{
\e^{i{\vekexp p}\cdot ({\vekexp u}_1-{\vekexp u}_2)} \hat K({\vek p})
\Delta \hat K({\vek p}) + \displaystyle{{1\over 2}} \big(
\e^{-i{\vekexp p}\cdot {\vekexp u}_2} \Delta_p [\e^{i{\vekexp p}\cdot 
{\vekexp u}_1}] \right. \\ & & 
\left. + \e^{i{\vekexp p}\cdot {\vekexp u}_1} \Delta_p 
[\e^{-i{\vekexp p}\cdot {\vekexp u}_2}] \big) +  \hat K({\vek p})
\nabla \hat K({\vek p}) \cdot \nabla_p \e^{i{\vekexp p}\cdot
({\vekexp u}_1-{\vekexp u}_2)} \right\} \\
\end{array} \eqno(A5)
$$
Integrating per partes once more the last term in (A5), we finally
arrive at
$$
\begin{array}{rcl}
f({\vek u}_1,{\vek u}_2) & = & \displaystyle{{1\over 2}}
(u_1^2 + u_2^2) \displaystyle{\int} \e^{i{\vekexp p}\cdot
({\vekexp u}_1-{\vekexp u}_2)} \hat K^2({\vek p}) \rd {\vek p} \\ & &
+ \displaystyle{\int} \e^{i{\vekexp p}\cdot ({\vekexp u}_1-
{\vekexp u}_2)} \vert \nabla \hat K({\vek p})\vert^2 \rd {\vek p} \\   
\end{array} \eqno(A6)
$$
With regard to the ``decoration'' relation (A2b), (A6) can be 
written in a more consistent form
$$
\begin{array}{rcl}
f({\vek u}_1,{\vek u}_2) & = & \displaystyle{{1\over 2}}
(u_1^2 + u_2^2)\displaystyle{{\partial K(\vert {\vek u}_1 - {\vek u}_2 \vert )
\over \partial n}} + \Phi(\vert {\vek u}_1 - {\vek u}_2 \vert )
\\
\Phi({\vek u}) & = & \displaystyle{\int} \e^{i{\vekexp p}\cdot {\vekexp u}} 
\vert \nabla \hat K({\vek p})\vert^2 \rd {\vek p} \\   
\end{array}\eqno(A7) 
$$
This equation admits a trivial generalization
$$
\begin{array}{rll}
&  \displaystyle{\int} \vert {\vek r} - {\vek u} \vert^2 
K(\vert {\vek u}_1 -{\vek r}\vert)
K(\vert {\vek r} -{\vek u}_2\vert) \rd {\vek r} & \\
& \quad \quad \quad = \displaystyle{{1\over 2}}
\left( \vert {\vek u}_1-{\vek u} \vert^2 + \vert {\vek u}_2- 
{\vek u}\vert^2 \right)
\displaystyle{{\partial K(\vert {\vek u}_1 - {\vek u}_2 \vert )
\over \partial n}} + \Phi(\vert {\vek u}_1 - {\vek u}_2 \vert ) & \\
\end{array}\eqno(A8)
$$

The double application of formula (A8) solves immediately the problem
of two root points generated on two different bonds:

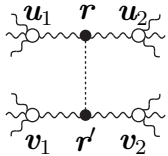
\begin{picture}(100,50)(-50,18)
    \Photon(10,10)(50,10){1}{8.5}
    \Photon(10,10)(0,10){1}{2}
    \Photon(10,10)(2,18){1}{2}
    \Photon(10,10)(2,2){1}{2}
    \Photon(50,10)(60,10){1}{2}
    \Photon(50,10)(58,18){1}{2}
    \Photon(50,10)(58,2){1}{2}
    \BCirc(10,10){2.5}
    \BCirc(50,10){2.5}
    \DashLine(30,10)(30,40){1}
    \Photon(10,40)(50,40){1}{8.5}
    \Photon(10,40)(0,40){1}{2}
    \Photon(10,40)(2,48){1}{2}
    \Photon(10,40)(2,32){1}{2}
    \Photon(50,40)(60,40){1}{2}
    \Photon(50,40)(58,48){1}{2}
    \Photon(50,40)(58,32){1}{2}    
    \Vertex(30,40){2.2}
    \Vertex(30,10){2.2}
    \BCirc(10,40){2.5} \BCirc(50,40){2.5}
    \Text(12,-1)[]{$\vek v_1$}
    \Text(30,1)[]{$\vek r'$}
    \Text(48,-1)[]{$\vek v_2$}
    \Text(12,48)[]{$\vek u_1$}
    \Text(30,49)[]{$\vek r$}
    \Text(48,48)[]{$\vek u_2$}
\end{picture}

\bigskip

\ni The final result reads
$$
\begin{array}{ll}
& \displaystyle{\int} \vert {\vek r} - {\vek r}'\vert^2
K(\vert {\vek u}_1-{\vek r} \vert) K(\vert {\vek r}-{\vek u}_2 \vert) 
K(\vert {\vek v}_1-{\vek r}' \vert) K(\vert {\vek r}'-{\vek v}_2 \vert) 
\rd {\vek r} \rd {\vek r}'
\\ & \quad \quad \quad
= \displaystyle{{1\over 4}} \left[ 
\vert {\vek u}_1 - {\vek v}_1 \vert^2 +
\vert {\vek u}_1 - {\vek v}_2 \vert^2 +
\vert {\vek u}_2 - {\vek v}_1 \vert^2 +
\vert {\vek u}_2 - {\vek v}_2 \vert^2  \right]
\\ & \quad \quad \quad \quad \times
\displaystyle{{\partial K(\vert {\vek u}_1 - {\vek u}_2 \vert)
\over \partial n}}
\displaystyle{{\partial K(\vert {\vek v}_1 - {\vek v}_2 \vert)
\over \partial n}}  
\\ & \quad \quad \quad \quad
+ \displaystyle{{\partial K(\vert {\vek u}_1 - {\vek u}_2 \vert)
\over \partial n}} \Phi(\vert {\vek v}_1 - {\vek v}_2 \vert)
+ \displaystyle{{\partial K(\vert {\vek v}_1 - {\vek v}_2 \vert)
\over \partial n}} \Phi(\vert {\vek u}_1 - {\vek u}_2 \vert)
\\
\end{array}\eqno(A9)
$$

When the two root points are generated on the same bond,

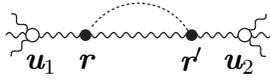
\begin{picture}(150,50)(-50,2)
    \Photon(10,10)(30,10){1}{4.5}
    \Photon(30,10)(70,10){1}{8.5}
    \Photon(70,10)(90,10){1}{4.5}
    \Photon(0,10)(10,10){1}{2}
    \Photon(10,10)(2,18){1}{2}
    \Photon(10,10)(2,2){1}{2}
    \Photon(90,10)(100,10){1}{2}
    \Photon(90,10)(98,18){1}{2}
    \Photon(90,10)(98,2){1}{2}
    \DashCArc(50,0)(22,22,143){1}
    \BCirc(10,10){2.5}
    \BCirc(90,10){2.5}
    \Vertex(30,10){2.2}
    \Vertex(70,10){2.2}
    \Text(13,-1)[]{$\vek u_1$}
    \Text(30,0)[]{$\vek r$}
    \Text(70,1)[]{$\vek r'$}
    \Text(88,-1)[]{$\vek u_2$}
\end{picture}

\medskip

\ni there holds
$$
\begin{array}{ll}
& \displaystyle{\int} 
K(\vert {\vek u}_1-{\vek r} \vert)
\vert {\vek r} - {\vek r}' \vert^2
K(\vert {\vek r}-{\vek r}' \vert)
K(\vert {\vek r}'-{\vek u}_2 \vert)
\\ & \quad \quad \quad \quad
= - 2 \pi \displaystyle{\int} \e^{i{\vekexp p}\cdot ({\vekexp u}_1
-{\vekexp u}_2)} \displaystyle{{\hat K}^2({\vek p}) 
} \Delta \hat K({\vek p}) \rd {\vek p}
\\ & \quad \quad \quad \quad
\equiv \Psi(\vert {\vek u}_1-{\vek u}_2 \vert) \\
\end{array}\eqno(A10)
$$

\vskip 2truecm

\ni {\bf APPENDIX B}
\medskip

In this part, we establish the transition from relation (\ref{47}) to 
Eq.(\ref{48}).
The origin of bond-dependent $F$-factors is irrelevant with the only 
proviso that the integrals exist.

Let us first consider the integral over infinite 2d space
$${1\over V} \int \vert {\vek r}_1 - {\vek r}_2 \vert^2 
f(\vert {\vek r}_1 - {\vek r}_2 \vert) \rd {\vek r}_1 \rd {\vek r}_2 =
\int r^2 f({\vek r}) \rd {\vek r} \eqno(B1)$$
written in the center-of-mass inertia; the supposed translational
invariance of function $f_{12} \equiv f(r_{12})$ is important.
The scaling transformation of coordinate ${\vek r} \to (1+\lambda)
{\vek r}$ does not alter the infinite boundary, and hence
$$\int r^2 f({\vek r}) \rd {\vek r} = (1+\lambda)^4 \int r^2
f[(1+\lambda){\vek r}] \rd {\vek r} \eqno(B2)$$
Expanding $(1+\lambda)^4=1+4\lambda+6\lambda^2+O(\lambda^3)$ and
$$f({\vek r} +\lambda {\vek r}) = f({\vek r}) + \lambda ({\vek r}
\cdot \nabla)f +{1\over 2}\lambda^2 ({\vek r}\cdot \nabla)^2 f
- {1\over 2} \lambda^2 ({\vek r}\cdot \nabla)f + O(\lambda^3)
\eqno(B3)$$
in Eq.(B2), the nullity of coefficients to the $\lambda$ and 
$\lambda^2$ powers implies, respectively,
$$
\begin{array}{rclr}
\hskip 2.3truecm 0 & = & \displaystyle{{1\over V} \int 
\vert 1-2 \vert^2 \left[ 4 f_{12} + (\bR_{12} f_{12}) \right] \rd 1 \rd 2  
} & \hskip 2truecm (B4a) \nonumber  \\
\hskip 2.3truecm 0 & = & \displaystyle{{1\over V} \int 
\vert 1-2 \vert^2 \left[ 6 f_{12} + {7\over 2} (\bR_{12} f_{12})+ 
{1\over 2} (\bR_{12}^2 f_{12}) \right] \rd 1 \rd 2 
} &\hskip 2truecm (B4b) \nonumber \\
\end{array}
$$
where we have adopted the operator notation given in (\ref{46}).
Note that the sum of two zeros $\{ {\rm Eq.}(B4a) \} /8 + 
\{ {\rm Eq.} (B4b)\} /4$,
$$0 = {1\over V} \int \vert 1-2 \vert^2 \left[
2f_{12} + (\bR_{12} f_{12}) +{1\over 8} (\bR_{12}^2 f_{12}) \right]
\rd 1 \rd 2 \eqno(B5)$$
with substitution $f_{12} = F_{12}$ corresponds to the rhs of (\ref{47})
for the simplest $N=2$ case.

Let us introduce the functions $g_{12}$ and $\bar g_{12}$ as follows
$$
\begin{array}{rclr}
\hskip 1.7truecm F_{12} g_{12} & = & \displaystyle{ \int \rd 3 \ldots \rd N 
\prod_{u,v=1 \atop (u<v)}^N F_{uv} } & \hskip 1.5truecm (B6) \\
\hskip 1.7truecm F_{12} {\bar g}_{12} & = & \displaystyle{ \int \rd 3 \ldots 
\rd N \prod_{u,v=1 \atop (u<v)}^N F_{uv} \sum_{i=3}^N 
\left[ {1\over F_{1i}}(\bR_{1i} F_{1i}) 
+ {1\over F_{2i}}(\bR_{2i} F_{2i})\right] } & \hskip 1.5truecm (B7) \\
\end{array}
$$
Both functions evidently possess the translational-invariance property.
The sequence of operations
$\{ {\rm Eq.}(B5)\ {\rm for}\ f_{12}=F_{12}g_{12} \} - \{ {\rm Eq.}
(B4a)\ {\rm for}\ f_{12}=F_{12}(\bR_{12} g_{12})\} /4 + \{ {\rm Eq.}
(B4a)\ {\rm for}\ f_{12}=F_{12}{\bar g}_{12} \} /8$ leads to
$$
\begin{array}{rcl}
0 & = & \displaystyle{ {1\over V} \int \rd 1 \rd 2 ~ \vert 1-2 \vert^2
\big[ 2F_{12}g_{12} + (\bR_{12}F_{12})g_{12} + {1\over 8} (\bR^2_{12}
F_{12}) g_{12} + {1\over 2} F_{12} {\bar g}_{12} } \\
& & \hskip 3.3truecm \displaystyle{ + {1\over 8} (\bR_{12} F_{12})
{\bar g}_{12} + {1\over 8} F_{12} (\bR_{12} {\bar g}_{12}) -{1\over 8}
F_{12} (\bR_{12}^2 g_{12}) \big] }
\end{array}\eqno(B8)
$$
\ni We take advantage of the flexibility in choosing operator
$\bR$ (\ref{46}), and evaluate $\bR^2_{12} g_{12}$ first by applying
$\bR_{12} = ({\vek r}_1 - {\vek r}_2)\cdot \nabla_1$ and then
by applying
$\bR_{12} = ({\vek r}_2 - {\vek r}_1)\cdot \nabla_2$,
with the result
$$
\begin{array}{rcl}
F_{12}(\bR_{12}^2 g_{12}) & = & F_{12} (\bR_{12} g_{12}) +
F_{12} (\bR_{12} {\bar g}_{12}) + \displaystyle{ \int \rd 3 \ldots \rd N 
\prod_{u<v} F_{uv} \sum_{i,j=3}^N {1\over F_{1i} F_{2j}}} \\
& & \times \big\{ \left[ ({\vek r}_2 -{\vek r}_i)\cdot \nabla_1 F_{1i}
\right] \left[ ({\vek r}_1 -{\vek r}_j)\cdot \nabla_2 F_{2j} \right]
-(\bR_{1i} F_{1i}) (\bR_{2j} F_{2j}) \big\}
\end{array}\eqno(B9)
$$
The substitution of (B9) into (B8) gives
$$
\begin{array}{rcl}
0 & = & \displaystyle{ {1\over V} \int \prod_i \rd i \prod_{u<v}F_{uv}
\vert 1-2 \vert^2} \\
& \times & \displaystyle{ \big\{ 2 + {1\over F_{12}}(\bR_{12}F_{12})
+ {1\over 8F_{12}}(\bR_{12}^2 F_{12})} \\
& + & \displaystyle{ {1\over 2} \sum_{i=3}^N \left[ {1\over F_{1i}}
(\bR_{1i}F_{1i})+{1\over F_{2i}}(\bR_{2i}F_{2i}) \right]} \\
& + & \displaystyle{{1\over 8} {1\over F_{12}}(\bR_{12} F_{12}) 
\sum_{i=3}^N \left[ {1\over F_{1i}}(\bR_{1i}F_{1i})+
{1\over F_{2i}}(\bR_{2i}F_{2i}) \right]}\\
& + & \displaystyle{ {1\over 8} \sum_{i,j=3}^N {1\over F_{1i} F_{2j}} 
(\bR_{1i} F_{1i}) (\bR_{2j} F_{2j}) } \\
& - & \displaystyle{ {1\over 16} \sum_{i=3}^N \left[
{1\over F_{1i}}({\vek r}_1 -{\vek r}_2)\cdot \nabla_1 F_{1i} +
{1\over F_{2i}}({\vek r}_2 -{\vek r}_1)\cdot \nabla_2 F_{2i} \right] 
} \\
& - & \displaystyle{{1\over 8} \sum_{i,j=3}^N {1\over F_{1i} F_{2j}}
\left[ ({\vek r}_2 -{\vek r}_i)\cdot \nabla_1 F_{1i} \right] 
\left[ ({\vek r}_1 -{\vek r}_j)\cdot \nabla_2 F_{2j} \right]
\big\}}
\end{array}\eqno(B10)
$$
Combining Eq.(B10) for each pair of skeleton vertices with Eq.(\ref{47}),
the latter reduces to relation (\ref{48}).


\end{document}